\documentclass[aps, prb, reprint]{revtex4-2}
\usepackage[english]{babel}
\usepackage{amsmath}
\usepackage{graphicx}
\usepackage{amssymb}
\usepackage{siunitx}

\usepackage{comment}
\usepackage{hyperref} 
\usepackage{soul} 
\providecommand\vect[1]{\boldsymbol{#1}}

\begin{document}

\title{Hopfions in screw chiral magnets}

\author{Sandra Chulliparambil Shaju}
\email[]{sandra.shaju@uni-due.de}
\affiliation{Faculty of Physics and Center for Nanointegration Duisburg-Essen (CENIDE), University of Duisburg-Essen, 47057 Duisburg, Germany}

\author{Maria Azhar}
\affiliation{Faculty of Physics and Center for Nanointegration Duisburg-Essen (CENIDE), University of Duisburg-Essen, 47057 Duisburg, Germany}

\author{Karin Everschor-Sitte}
\affiliation{Faculty of Physics and Center for Nanointegration Duisburg-Essen (CENIDE), University of Duisburg-Essen, 47057 Duisburg, Germany}

\date{\today}

\begin{abstract}
Three-dimensional topological spin textures have attracted growing interest due to their rich geometry and potential for functional magnetic phenomena. 
In this work, we propose the concept of symmetry-transforming magnetic models as a novel route to generate and stabilize complex three-dimensional textures in an arbitrary magnetic background. Using this framework, we predict a screw chiral magnet model that stabilizes magnetic Hopfions and other three-dimensional magnetic textures within a ferromagnetic background. We show that the resulting solitons display distinctive physical properties, including unconventional Goldstone modes. Our results establish continuous symmetry transformations as a general strategy for uncovering new classes of magnetic solitons with unique dynamical signatures.
\end{abstract}


\maketitle 

\section{Introduction}
Topological spin textures in three dimensions (3D) have emerged as a vibrant frontier in modern magnetism, driven by both fundamental interest in their nontrivial geometry and the prospect of exploiting their robustness and dynamics in future spintronic applications. Beyond two-dimensional solitons such as skyrmions~\cite{Bogdanov1989, muhlbauer_2009, Fert_2017, everschor_2018, Back_2020, Mishra2025}, 3D textures~\cite{BALAKRISHNAN2023, Gubbiotti2025} -- including Hopfions~\cite{Sutcliffe_2017, kent_2021,Liu_2018}, torons~\cite{Leonov_2018, Liu_2018, Muller_2020}, chiral bobbers~\cite{zheng_2018, Ran2021, Rybakov_2015},
heliknotons~\cite{Voinescu2020,Kuchkin_2023, Li2026, Zheng_2025}, screw dislocations~\cite{azhar_2022, Maria_2024, zheng_2023}, 
and other knotted or twisted field structures -- enrich the magnetic landscape by hosting complex topology characterized by integer-valued invariants~\cite{everschor_2018, Whitehead_1947, azhar_2022, Maria_2024} and by supporting unique dynamical modes  \cite{Voinescu2020, Kravchuk2023}. 
Understanding the mechanisms that enable their formation and stability in real materials is therefore of central importance.

Chiral magnets provide a natural setting for twisted spin textures, as Dzyaloshinskii–Moriya interactions (DMI) arising from broken inversion symmetry promote spatially modulated states~\cite{Dzyaloshinskii1958, Moriya1960}. In two dimensions (2D), textures such as domain walls, helices, and skyrmions are well established in these systems. More recently, experimental studies have also uncovered signatures of three-dimensional magnetic textures embedded within a helical magnetic state~\cite{zheng_2023, Voinescu2020, Kuchkin_2023, azhar_2022, zheng_2023, Maria_2024, Zheng_2025, Li2026}.
While helices and skyrmion crystals arise naturally from the intrinsic twisting favored by DMI, isolated 3D solitons in a uniform ferromagnetic background lack an inherent energetic mechanism to sustain their spatial variation and have so far required finely tuned combinations of higher-order interaction terms~\cite{Bogolubsky_1988, Sutcliffe_2017, Rybakov_2022}.

In this work, we introduce the concept of symmetry-transforming magnetic models. In particular, we show that screw-transforming a basic chiral magnet model generates a novel chiral model in which magnetic Hopfions and more intricate magnetic 3D magnetic textures embedded in a ferromagnetic background are metastable.
Importantly, the resulting Hopfions, obtained by screw-transforming a heliknoton stabilized in a helimagnetic background, exhibit distinct physical characteristics, including unconventional Goldstone modes associated with the continuous symmetries of their screw-twisted structure. These features distinguish them from Hopfionic states previously discussed in helical or frustrated magnetic backgrounds and point toward unique dynamical signatures that may be observable experimentally.

\section{Screw Chiral magnet and (de-)spiralization}
We introduce a screw chiral magnet micromagnetic model that stabilizes a plethora of 3D magnetic textures in a ferromagnetic background, including Hopfions. Its energy functional is
\begin{equation}
\label{eq: despiralized energy functional}
\begin{aligned}
 \mathcal{E} &= \mathcal{A} \: (\vect{\nabla}\vect{m})^2   
    + \mathcal{D} \: \sin{\varphi} \: \left(\mathcal{L}_{zx}^x + \mathcal{L}_{zy}^y \right) \\
    &+ \mathcal{D} \: \cos{\varphi} \: \left( \mathcal{L}_{zy}^x + \mathcal{L}_{xz}^y \right)
    +\frac{\mathcal{D}^2}{4\mathcal{A}} \: m_z^2 
\end{aligned}
\end{equation}
where $\vect{m}$ denotes the normalized magnetization vector field, $\mathcal{A}>0$ is the exchange stiffness, $\mathcal{L}^i_{jk}= m_j \partial_i m_k - m_k \partial_i m_j$ denote the Lifshitz invariants and $\mathcal{D}$ the maximal Dzyaloshinskii–Moriya interaction (DMI) strength. The angle 
\begin{equation}
\varphi=\mathcal{D}z/2\mathcal{A}
\label{eq:varphi}
\end{equation}
leads to a spatially-modulated interaction strength such that the coefficients of the Lifshitz terms vary periodically along $z$ with a screw symmetry: 
the Néel-type DMI $(\mathcal{L}_{zx}^x + \mathcal{L}_{zy}^y)$ varies as $\sin{\varphi}$, while the Bloch-type DMI $(\mathcal{L}_{zy}^x + \mathcal{L}_{xz}^y)$ varies as $\cos{\varphi}$.  
The interplay of the exchange and DMI terms determines the screw pitch $L_D=4\pi\mathcal{A}/\mathcal{D}$ 
as a natural length scale of the screw chiral magnet model.
In addition, Eq.~\eqref{eq: despiralized energy functional} includes an easy-plane anisotropy term of strength $\mathcal{D}^2/4\mathcal{A}$, which favors an in-plane ferromagnetic ground state.

\begin{figure}[htbp]
    \centering
  \includegraphics[width=\linewidth]{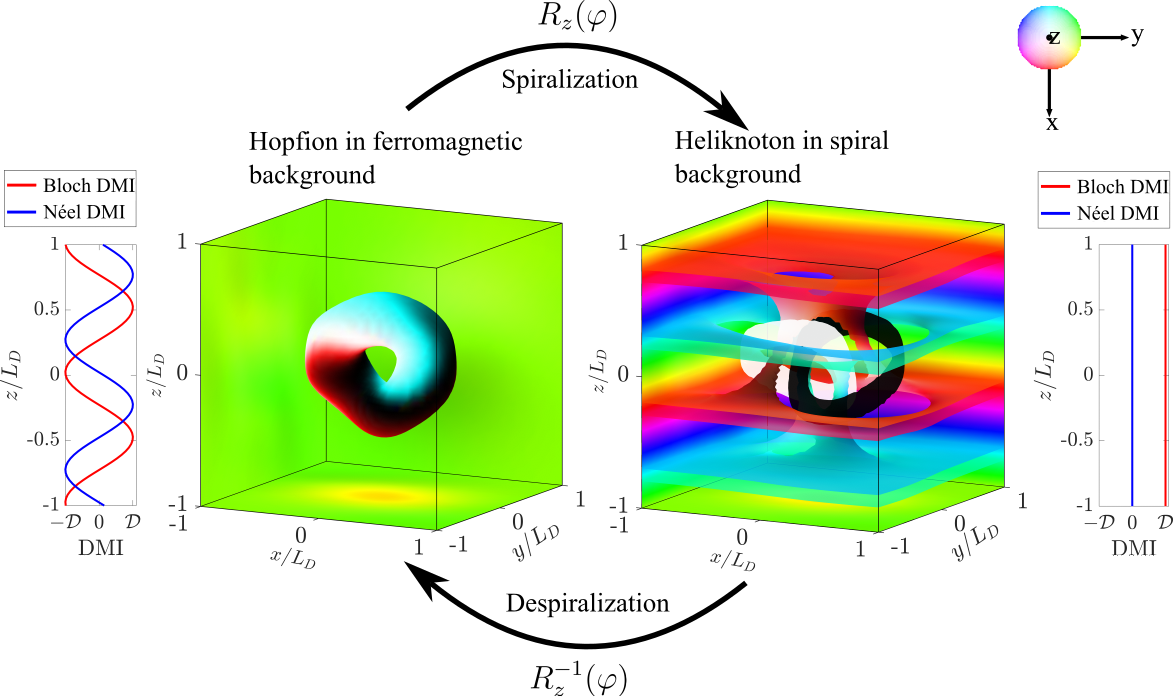}
    \caption{
    Spiralization and despiralization of magnetic textures and their corresponding models. 
    Left: a Hopfion stabilized in the ferromagnetic ground state of the screw-symmetric model (Eq.~\eqref{eq: despiralized energy functional}).
    Right: a heliknoton embedded in the spiral ground state of a conventional chiral-magnet model.
    These two textures can be transformed into each other via (de-)spiralization, and the associated magnetic models indicated on the outer panels by highlighting their DMI interactions are related by the same screw transformation. Displayed are the isosurfaces of $m_y=0$ and the magnetization along edges of the cubic sample of size $2 L_D$. The colormap indicates the magnetization direction, as defined in the inset.}
    \label{fig:four_isosurface}
\end{figure}

We constructed the screw chiral magnet model starting from a heliknoton \cite{Voinescu2020} that is stable in the bulk chiral magnet model $\mathcal{E}_{\mathrm{CM}}[\vect{m}] =\: \mathcal{A} \: (\vect{\nabla} \vect{m})^2 \: + \: \mathcal{D} \: \vect{m} \cdot (\vect{\nabla} \times \vect{m})$.
Despiralizing the heliknoton eliminates the helical modulation and transforms the heliknoton into a Hopfion embedded in a ferromagnetic background, see Fig.~\ref{fig:four_isosurface}.
This despiralization procedure can be described by performing a coordinate-dependent inverse spin rotation $R^{-1}_z(\varphi)$ by the angle $\varphi$ defined in Eq.~\eqref{eq:varphi} around the direction of the spiral wave-vector $\vect{q}$ (here taken along $z$), i.e.\
 $\vect{m}_\mathrm{Hopfion}= R^{-1}_z(\varphi)\vect{m}_\mathrm{heliknoton}$, where the screw transformation $R_z(\varphi)$ acts as
$R_z(\varphi)\vect{m}=
(m_x\cos\varphi-m_y\sin\varphi,\  
m_x\sin\varphi+m_y\cos\varphi, \ 
 m_z)$.
Applying this despiralization concept directly to the bulk chiral magnet model yields Eq.~\eqref{eq: despiralized energy functional} up to an additive constant,
$\mathcal{E}[\vect{m}]=\mathcal{E}[R^{-1}R\vect{m}]=\mathcal{E}[R^{-1}\vect{\tilde{m}}]=\mathcal{E}_{CM}[\vect{\tilde{m}}] +$ $\mathcal{D}^2/4\mathcal{A}$,
where $\vect{\tilde{m}}=R\vect{m}$~\footnote{
Simplified versions of this approach were proposed in Refs.~\cite{BorisovIzyumov1984, Zarzuela2020} for lower dimensional magnetic textures.}. 

While the bulk chiral-magnet model possesses full $SU(2)$ symmetry under simultaneous rotations in spin and real space, 
the symmetry of the screw chiral magnet reduces to a continuous $U(1)$ symmetry represented by the joint transformation $(\mathbf{r},\mathbf{m}) \mapsto (R_z(\theta)\,\mathbf{r},\, R_z(\theta)\,\mathbf{m})$ with arbitrary rotational angle $\theta$.

The ground states of the screw chiral magnet $m^{gs}$ are obtained by despiralizing the ground states of the bulk chiral magnet $\vect{m}_{CM}^{gs}$, i.e.\ $\vect{m}^{gs}=R^{-1}_z(\varphi)\vect{m}_{CM}^{gs}$. 
Due to the rotational symmetry, the bulk chiral-magnet model admits an $SU(2)$-symmetric manifold of degenerate ground states consisting of spirals with pitch $L_D$ and arbitrary orientation of the spiral wavevector $\vect{q}$. 
Upon despiralization, i.e.\ applying the rotation along the $\vect{q}$ vector direction, these transform into uniform in-plane ferromagnetic (FM) states~\footnote{Please note that one can, in principle, apply the rotation also along a different direction than the spiral axis.  This then leads to doubly modulated helical states with an example shown in App.~\ref{sec:othermagn} in Fig.~\ref{fig:2q spiral}}.

Analogously, metastable structures can also be derived in the screw–chiral-magnet model. Topological defects that are metastable in the conventional chiral-magnet model have direct counterparts in the screw-transformed formulation, Eq.~\eqref{eq: despiralized energy functional}, and can be computed by despiralization.
In the next sections, we first address magnetic Hopfions in the screw transformed model (the counterparts of heliknotons in the bulk chiral model), their excitations, and then further topologically non-trivial textures that arise in the screw chiral model.

\section{Hopfions in screw-chiral magnets}
A screw-chiral magnet provides an ideal setting for realizing magnetic Hopfions due to the coexistence of Néel- to Bloch-type interactions.
The screw symmetry inherent in Eq.~\eqref{eq: despiralized energy functional} results in a manifold of Hopfions with different background orientations that can be stabilized.
(see also Sec.~\ref{sec:Goldstone} for a discussion of the associated Goldstone modes).

Using micromagnetic simulations, we computed the stable Hopfion configurations stabilized by Eq.~\eqref{eq: despiralized energy functional}, see Methods for details. These solutions are generally distorted compared to perfect toroidal Hopfions, as illustrated in the left panel of Fig.~\ref{fig:four_isosurface}. In the example shown, the Hopfion ring lies predominantly in the 
$xz$ plane and is embedded in a ferromagnetic background with orientation $\vect{m}^{\mathrm{bg}}=(0,1,0)$.
The Hopfion centreline, defined by the isoline $\vect{m}=-\vect{e}_y$ is not circular but has an oscillatory displacement above and below the Hopfion's midplane (defined as the plane perpendicular to the Hopfion’s global axis of symmetry that passes through the center of mass of the centreline). 
This out-of-plane displacement exhibits a twofold periodicity along the toroidal coordinate. When viewed along the $y$-axis, the centreline exhibits a four-fold distortion (see App.~\ref{app:details}).

\section{Hopfion Symmetries and Their Associated Zero Modes}
\label{sec:Goldstone}

In this section, we identify the continuous symmetry transformations under which the Hopfion has the same energy, i.e.\ those that leave the energy functional in Eq.~\eqref{eq: despiralized energy functional} invariant.
These symmetries, denoted by screw and gyration as well as their combined transformation denoted by swirl, give rise to nontrivial Goldstone modes, summarized in Table~\ref{table:zero_modes}.

\begin{figure}[tbp]
    \centering
\includegraphics[width=0.97\linewidth]{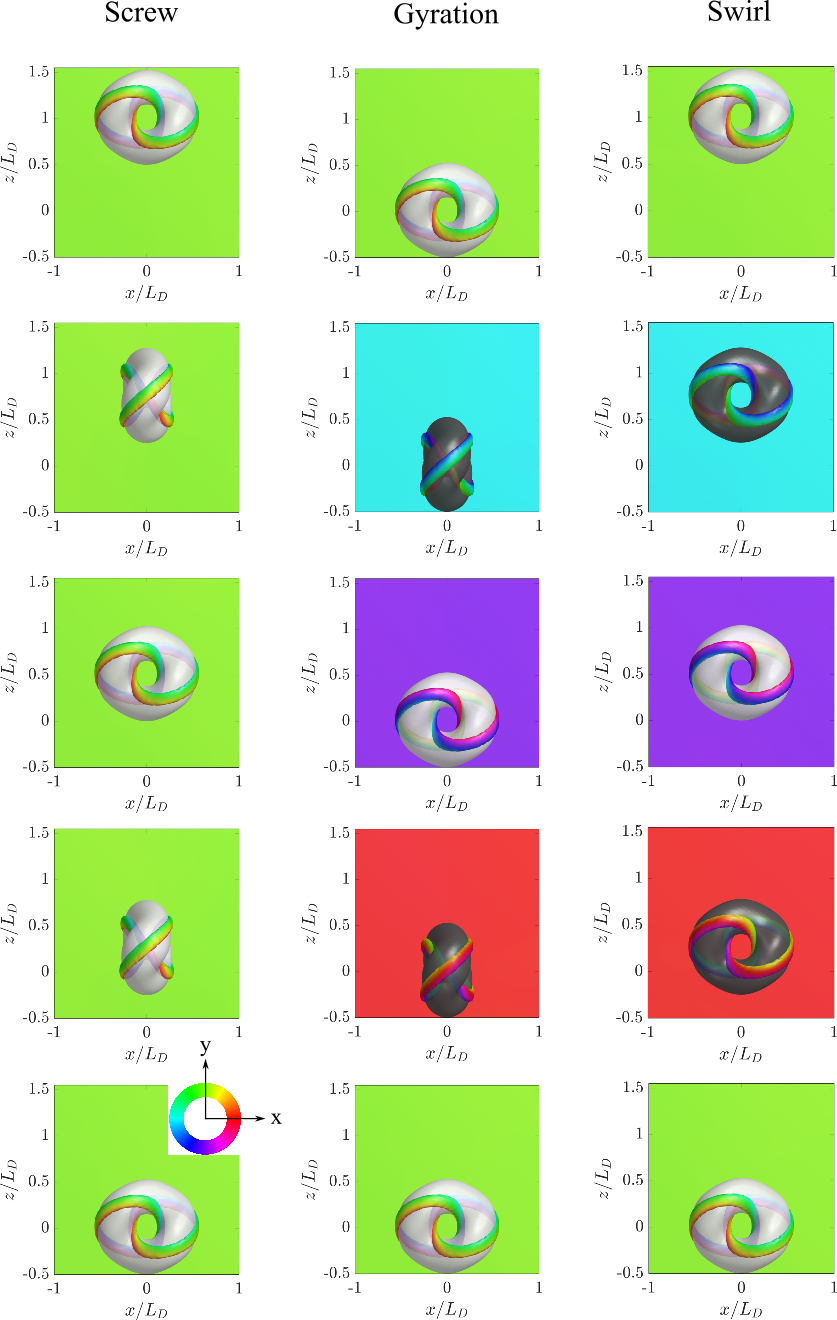}
    \caption{
    Overview of the Goldstone modes of the Hopfion: screw (left), involving simultaneous changes of $z_0$ and $\phi$; gyration (middle), involving only a change of $\phi$; and swirl (right), involving only a change of $z_0$, which is a combination of screw and gyration motion. Shown are representative snapshots along each zero-mode trajectory. The background magnetization lies in the \(xy\)-plane, and its direction is indicated by the HSV colormap (see inset in the lower-left panel). In each image, either the \(m_y = 0\) or \(m_x = 0\) isosurface (white and black, respectively) is displayed, highlighting the characteristic toroidal structure. The \(m_z = \pm 0.99\) isosurfaces are shown in all panels and are colored according to the in-plane magnetization component. 
    }
     \label{fig:zero_modes}
\end{figure}

\subsection*{Mode I: Screw}
The screw chiral magnet energy, Eq.\eqref{eq: despiralized energy functional}, is invariant under the screw transformation  
\begin{equation}
\textbf{Screw:}\quad  (\vect{r},\vect{m}) \;\mapsto\; \bigl(R_z(\phi)\,\vect{r}+z_0(\phi) \mathbf{e}_z,\; \vect{m}\bigr),
\end{equation}
which consists of an identity operation in spin space combined with a rotation of the coordinate frame about the 
$z$-axis by an angle $\phi$ and a corresponding translation by $z_0(\phi)\vect{e}_z$, where
$z_0(\phi)= L_D \phi/(2\pi)$.

As a continuous symmetry, this transformation generates a Goldstone mode describing a screw-like propagation of the Hopfion along the \(z\)-direction. During this motion, the Hopfion midplane rotates about the \(z\)-axis while the Hopfion center translates along \(+z\) for \(z_0 < 0\), whereas the uniform background magnetization remains fixed. This collective motion is illustrated in the left column of Fig.~\ref{fig:zero_modes}.

\begin{table}[tbp]
\centering
\renewcommand{\arraystretch}{1.15}
\begin{tabular}{llccl}
\hline
\multicolumn{2}{l}{Mode} &\ $z$-shift\ & \ midplane rot.\ &  bg behaviour \\
\hline
I &Screw  & $+z_0$ & $+\phi$ &  
bg fixed \\
II &Gyration & none   & $+\phi$ & bg rotates \\
I\&II &Swirl & $+z_0$ & none   & 
bg rotates \\
\hline
\end{tabular}
\caption{Summary of the Hopfion’s zero modes and their effects on its translation ($z$-shift), midplane rotation, and background (bg) magnetization.}
\label{table:zero_modes}
\end{table}

\subsection*{Mode II: Gyration}
The energy functional is also invariant under the gyration transformation 
\begin{equation}
  \textbf{Gyration:}\quad    (\mathbf{r},\mathbf{m}) \;\mapsto\; \bigl(R_z(\phi)\mathbf{r},\;R_z(\phi) \mathbf{m}\bigr)
    \end{equation}
which corresponds to simultaneous rotations of both real space and spin space about the $z$-axis. The associated Goldstone mode 
describes a rotation of the Hopfion midplane together with the background magnetization around the $z$-axis without translation, see middle column of Fig.~\ref{fig:zero_modes} for $\phi<0$. This rotation of the background magnetization is an analog of the Archimedean spiral mode discussed in Ref.~\cite{delSer2021}.

\subsection*{Combining Mode I and II: Swirl}
A simultaneous excitation of the screw and gyration modes gives rise to a
\begin{equation}    
\textbf{Swirl:} \quad(\mathbf{r},\mathbf{m}) \;\mapsto\; \bigl(\mathbf{r}+z_0(\phi)\mathbf{e}_z,\;R_z(-\phi) \mathbf{m}\bigr),
\end{equation}
which likewise leaves the energy functional invariant.
This mode couples a translation along the $z$ axis to a compensating rotation of the magnetization about the same axis, where again $z_0(\phi)= L_D \phi/(2\pi)$.
The corresponding Goldstone mode describes a motion in which the Hopfion is displaced along the 
$z$-direction without a rotation of its midplane, while its preimages undergo an internal swirl and the background magnetization rotates within the 
$xy$-plane. This mode is illustrated in the right column of Fig.~\ref{fig:zero_modes}.

\section{Topological Implications of Despiralization}
When a three-dimensional magnetic texture is despiralized, 
its topological characteristics, such as
linking numbers and the Hopf index\footnote{Note that for periodic structures the Hopf index is defined per period.}, may change. 
In the following, we analyze these changes using the formalism introduced in Ref.~\cite{Maria_2024}, which is particularly well suited for this purpose.

Within this approach, the texture is decomposed into flux tubes
of the emergent magnetic field $\vect{F}$, defined as
$F^k = \epsilon^{ijk}\,\vect{m} \cdot (\partial_i \vect{m} \times \partial_j \vect{m})/8\pi$, where $\epsilon^{ijk}$ is the Levi-Civita symbol and $i,j,k \in \{x,y,z\}$.
Each flux tube carries a flux
$\Phi_i=\left|\int_{\vect{a}_i}\mathrm{d}\vect{a}\cdot\vect{F}\right|$, where the integration is performed over the cross-sectional area $\mathbf{a}_i$ of the corresponding flux tube.
In this representation, the Hopf index reads~\cite{Maria_2024}
\begin{equation}
H=\sum_{i=1}^{N_\Phi} L_{ii}\Phi_i^2
+ \sum_{\substack{i,j=1\\ j\neq i}}^{N_\Phi} L_{ij}\Phi_i\Phi_j,
\label{eq:Hopf_decomposed}
\end{equation}
where $N_\Phi$ is the number of flux tubes. $L_{ii}$ and $L_{ij}$ represent the oriented self-linking number of flux tube $i$ and the oriented inter-linking number between flux tubes $i$ and $j$, respectively.

\begin{figure*}[t]
    \centering
    \includegraphics[width=0.9\linewidth]{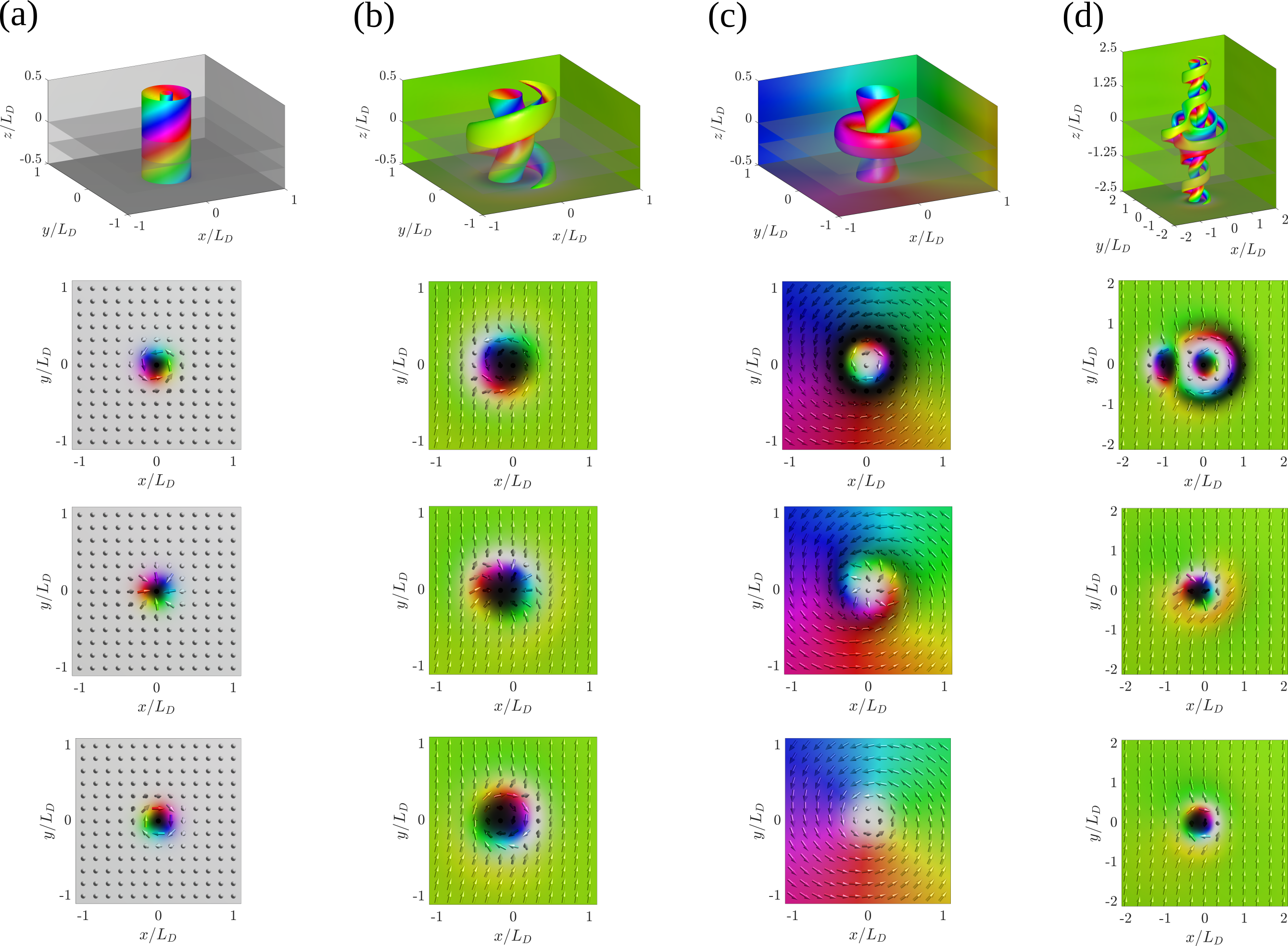}
    \caption{Examples of stable magnetization configurations in the screw chiral magnet. Highlighted are isosurfaces of $m_z = \pm 0.75$ for the four three-dimensional spin textures: (a) a twisted skyrmion tube, (b) a twisted in-plane skyrmion tube, (c) a despiralized Twiston, and (d) a despiralized Hopfion configuration surrounding a skyrmion string. For configurations (a)--(c) a full period along the $z$-axis is shown and the subfigures depict $xy$-plane cross-sections at $z = 0$, $z = -L_D/4$, and $z = -L_D/2$. Configuration (d) displays five periods of the skyrmion string with a Hopfion ring located near the $z = 0$ plane. The subfigures for (d) show the $xy$-plane at $z = 0$, $z = -5L_D/4$, and $z = -5L_D/2$.    }
    \label{fig:other_textures}
\end{figure*}

Starting from a magnetic state with a spiral background, the flux tubes of $\vect{F}$ can be classified into those entirely confined within the sample and those intersecting the boundary.
Upon despiralization, bulk-confined flux tubes retain their self-linking numbers, whereas boundary-intersecting flux tubes acquire modified self-linking numbers determined by the despiralization.
The inter-linking between distinct flux tubes, irrespective of their class, remains unchanged.
Consequently, the despiralization process also alters the Hopf index.

\begin{figure}[tb]
    \centering
    \includegraphics[width=\linewidth]{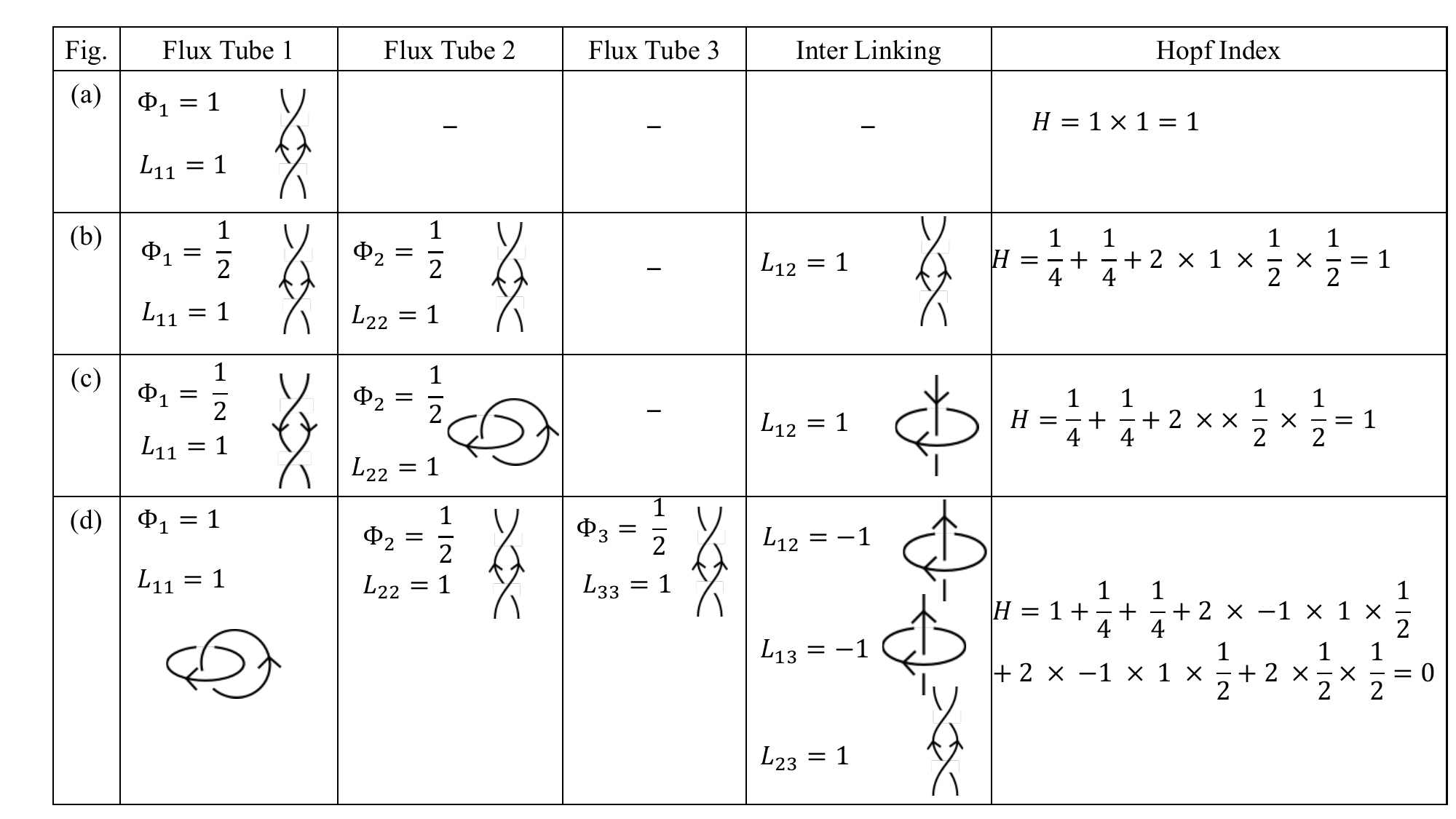}
    \caption{Summary of Linkings and Hopf index calculation for the spin textures shown in Fig.~\ref{fig:other_textures}. 
    For (a)–(d), the Hopf index $H$ is calculated for one period along $z$. For (d) $H$ is evaluated for the region $-0.5<z<0.5$, see Fig.~\ref{fig:other_textures}(d).}
    \label{fig:table_L_H}
\end{figure}

To illustrate this behavior, we present a topological characterization of representative three-dimensional textures in the bulk chiral magnet model and their despiralized counterparts in the screw chiral magnet.
The newly identified magnetic textures of the screw chiral magnet are shown in Fig.~\ref{fig:other_textures}, while their corresponding Hopf indices are summarized in Fig.~\ref{fig:table_L_H}.
For completeness and comparison, the previously reported magnetic textures of the bulk chiral magnet model are shown in App.~ Fig.~\ref{fig:other_textures_bcm}.

\paragraph{Skrymion tube $\rightarrow$ twisted skyrmion tube:}
In a sufficiently strong magnetic field applied along the $z$ direction, the bulk chiral magnet stabilizes a skyrmion tube (see App.\ Fig.~\ref{fig:other_textures_bcm}(a)).
This state can be described by a single emergent flux tube without self-linking~\cite{Maria_2024}, corresponding to a vanishing Hopf index, $H=0$.
Upon despiralization, the skyrmion tube acquires a helicity rotation, as illustrated in Fig.~\ref{fig:other_textures}(a).
The resulting texture is characterized by a single flux tube with self-linking number one per period, and thus has a Hopf index $H=1$ per period~\cite{Knapman_2024}.

\paragraph{In-plane skyrmion tube in spiral background $\rightarrow$ twisted in-plane skyrmion tube:}
The bulk chiral magnet model stabilizes an in-plane skyrmion tube embedded within a spiral background~\cite{Leonov_2016} (see App.\ Fig.~\ref{fig:other_textures_bcm}(b)). 
This texture consists of an antimeron tube carrying flux $\Phi=1/2$ with self-linking number $2$ per period, winding around a central meron tube without self-linking.
The inter-linking of the two tubes is $L_{12}=1$.
As a result, the Hopf index of the in-plane skyrmion tube in the spiral background is
$H=2*(1/2)^2 + 2*1*(1/2)*(1/2)=1$ 
per period $L_D$ along the $z$ direction~\cite{Maria_2024}.

Upon despiralization, the texture transforms into a twisted in-plane skyrmion tube in a uniform background, as shown in Fig.~\ref{fig:other_textures}(b).
In this process, the central meron tube acquires a self-linking number of $1$, while the surrounding antimeron tube reduces its self-linking to $1$.
The resulting configuration therefore, has the Hopf index
$H = (1/2)^2 + (1/2)^2 + 2*(1/2)^2=1$.

\paragraph{Twiston $\rightarrow$ despiralised Twiston:}

A Twiston~\cite{azhar_2022,  Maria_2024} is a screw dislocation composed of two flux tubes (see App.\ Fig.~\ref{fig:other_textures_bcm}(c)). 
It consists of a central vertical meron tube without self-linking, surrounded by a toroidal meron tube with self-linking number $1$.
The inter-linking between the two tubes is $1$.
Accordingly, the Hopf index of the Twiston is $H=1*(1/2)^2+ 2*1*(1/2)*(1/2) = 3/4$. 

Upon despiralization, the inner meron tube acquires a finite self-linking number, as shown in Fig.~\ref{fig:other_textures}(c).
The resulting despiralized Twiston \cite{kasai2025}, therefore, has a Hopf index
$H = (1/2)^2 + (1/2)^2 + 2*1*(1/2)*(1/2)= 1$.

\paragraph{Hopfion configuration around a skyrmion tube $\rightarrow$ despiralized Hopfion around a skyrmion tube}
A Hopfion configuration surrounding a skyrmion tube in a spiral background has been reported in Ref.~\cite{zheng_2023} (see App.~Fig.~\ref{fig:other_textures_bcm}(d)).
The texture spans five spiral periods and admits a decomposition into three distinct flux tubes:
a confined, torus-like Hopfion tube with self-linking number $L_{11}=1$, a central vertical meron tube with $L_{22}=0$, and a surrounding vertical antimeron tube with $L_{33}=2$.
The Hopfion tube is interlinked with both the meron and antimeron tubes, with linking numbers $L_{12}=-1$ and $L_{13}=-1$, respectively, while the meron and antimeron tubes are interlinked with $L_{23}=1$.
Within the spiral period containing the central Hopfion ring, this Hopf index contribution evaluates to
 $1+2*(1/2)^2-2*(1/2)-2*(1/2)-2*(1/2)*(1/2)=-1$.
 Each of the remaining four periods, which contain only the meron and antimeron tubes, contributes $2*(1/2)^2+2*(1/2)*(1/2)=1$.
 Consequently, the total Hopf index of the despiralized texture across the five screw pitches is $H=-1+ 4*1 = 3$.

A despiralized version of this configuration with a uniform magnetic background is shown in Fig.~\ref{fig:other_textures}(d).
It also spans five screw pitches and admits a decomposition into three distinct flux tubes:
a confined, torus-like Hopfion tube with unmodified self-linking number $L_{11}=1$, a meron tube with increased linking number $L_{22}=1$, and an antimeron tube with decreased linking number $L_{33}=1$.
The interlinking numbers remain unchanged.
Within the screw pitch containing the central Hopfion ring, this Hopf index contribution evaluates to
  $1 +(1/2)^2 + (1/2)^2 - 2 * \frac{1}{2} - 2 * (1/2) + 2 (1/2)*(1/2)= 0 $. 
Each of the remaining four periods, which contain only the meron and antimeron tubes, contributes $1$.
Consequently, the total Hopf index of the despiralized texture across the five screw pitches is $H=0+4*1=4$.

\section{Discussion and Conclusion}
In this work, we have introduced screw-transformed magnetic models as a general framework for stabilizing and controlling three-dimensional topological spin textures in arbitrary background states.
By exploiting continuous symmetry transformations that intertwine spatial translations with spin rotations, we extended the landscape of realizable three-dimensional magnetic solitons beyond those supported by conventional magnetic models.

Developing a screw chiral magnet model with spatially modulated Dzyaloshinskii--Moriya interactions, we predicted that Hopfions and a variety of other non-trivial three-dimensional magnetic textures can be stabilized within a uniform ferromagnetic background.
Beyond static stability, the screw-transformed nature of the magnetic textures gives rise to distinctive dynamical properties.
In particular, Hopfions in screw chiral magnets exhibit unconventional Goldstone modes that couple spatial motion to spin rotations, directly reflecting the underlying continuous screw symmetry of the model.
We expect these modes to be excitable by rotating magnetic fields in the 
$xy$ plane or by spatially modulated drives. Their chiral nature implies a natural coupling to circularly polarized light, enabling optical excitation at suitable frequencies and wave vectors. More generally, structured light carrying orbital angular momentum provides a promising route for selective control of these collective modes.

Our topological analysis shows that symmetry transformations modify not only the energy landscape but also topological invariants. While self-linking and mutual linking of internal emergent-field flux tubes are preserved under despiralization, self-linking associated with boundary-intersecting tubes changes, leading to a modified Hopf index and a clear geometric picture of how three-dimensional topology evolves under continuous symmetry transformations.

In summary, screw chiral magnets constitute a fertile playground for discovering, stabilizing, and manipulating three-dimensional magnetic solitons with tailored static and dynamical properties.
More broadly, our results establish continuous symmetry transformations as a general design principle for uncovering new classes of magnetic textures with unique topological and dynamical signatures.

\section{Acknowledgments}
We acknowledge fruitful discussions with Niklas Oettgen, Jacob Mankenberg, and Volodymyr Kravchuk. This work was supported by the German Research Foundation (DFG) under Project No.~505561633 as part of the TOROID project, co-funded by the French National Research Agency (ANR) under Contract No.~ANR-22-CE92-0032. Furthermore, we acknowledge funding from the DFG through Project No.~278162697 (SFB~1242, project~B10). M.~A.~acknowledges funding from the TORUS project within the UDE Postdoc Seed Funding program. S.~C.~S.~acknowledges support from the Studienstiftung des Deutschen Volkes.

AI-assisted language editing was used.

\newpage

\textcolor{white}{.}

\newpage

\appendix

\begin{center}
    {\Large \bfseries Supplementary Materials}
\end{center}

\section{Micromagnetic Simulations and Analysis}
Micromagnetic simulations were performed using the energy functional in Eq.~\eqref{eq: despiralized energy functional}, implemented in a custom package in \textsc{MuMax3}~\cite{Vansteenkiste2014} with the following material parameters: exchange stiffness $\mathcal{A} = \SI{0.4e-12}{\joule\per\meter}$, Dzyaloshinskii–Moriya interaction (DMI) constant $\mathcal{D} = \SI{0.28e-3}{\joule\per\square\meter}$, and saturation magnetization $\mathcal{M}_S = \SI{0.163e6}{\ampere\per\meter}$. For the results shown in the main text, the simulation grid size is $(N_x, N_y, N_z) = (120, 120, 120)$ with a uniform cell size of $\SI{0.75}{\nano\meter}$, corresponding to a cubic box of side length $5L_D$. 
Only for Fig.~\ref{fig:4distortion} of App.~\ref{app:details} we used a larger grid size of 
$(N_x, N_y, N_z) = (180, 180, 180)$ with a finer uniform cell size of $\SI{0.50}{\nano\meter}$.
Periodic boundary conditions were applied in all directions.

\section{Details of Hopfion analysis in the screw chiral magnet}
\label{app:details}

\begin{figure}[b]
    \centering
    \includegraphics[width=0.6\linewidth]{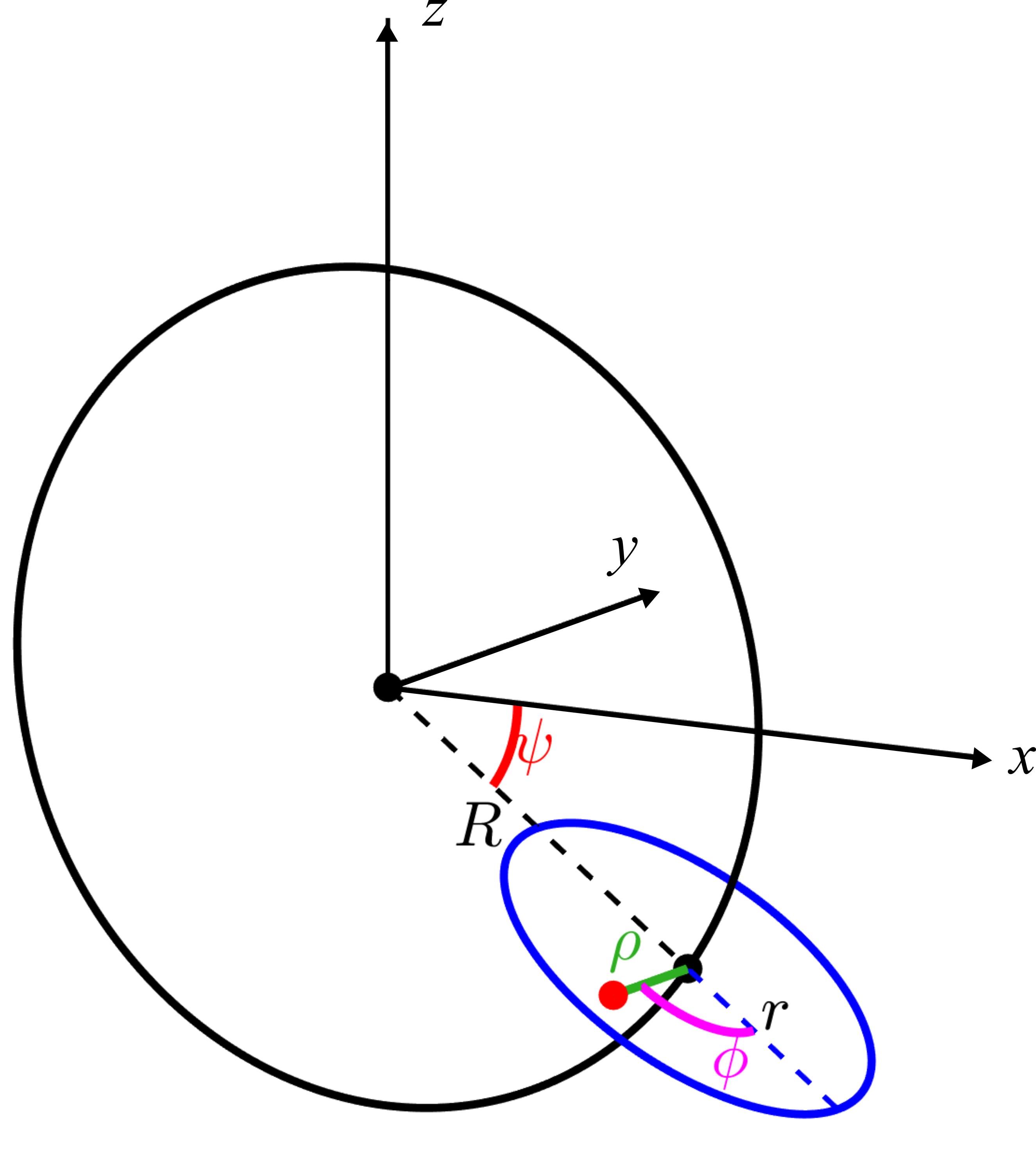}
    \caption{Schematic illustration of the Hopfion ansatz described by Eqs.~\eqref{eq:Hopfion_ansatz} to \eqref{eq:Hopfion_ansatz2}. 
The $\mathbf{m} = -\mathbf{e}_y$ isoline lies in the $xz$-plane on a circle of major radius $R$ (black circle). 
The angle $\psi$ locates the minor circle of radius $r$ (blue circle), which lies in a plane orthogonal to the tangent of the major circle. 
The local coordinates $(\rho,\phi)$ parametrize points within the minor circle. 
}
    \label{fig:Hopfion-ansatz}
\end{figure}

\begin{figure*}[ht]
    \centering
   \includegraphics[width=0.9\linewidth]{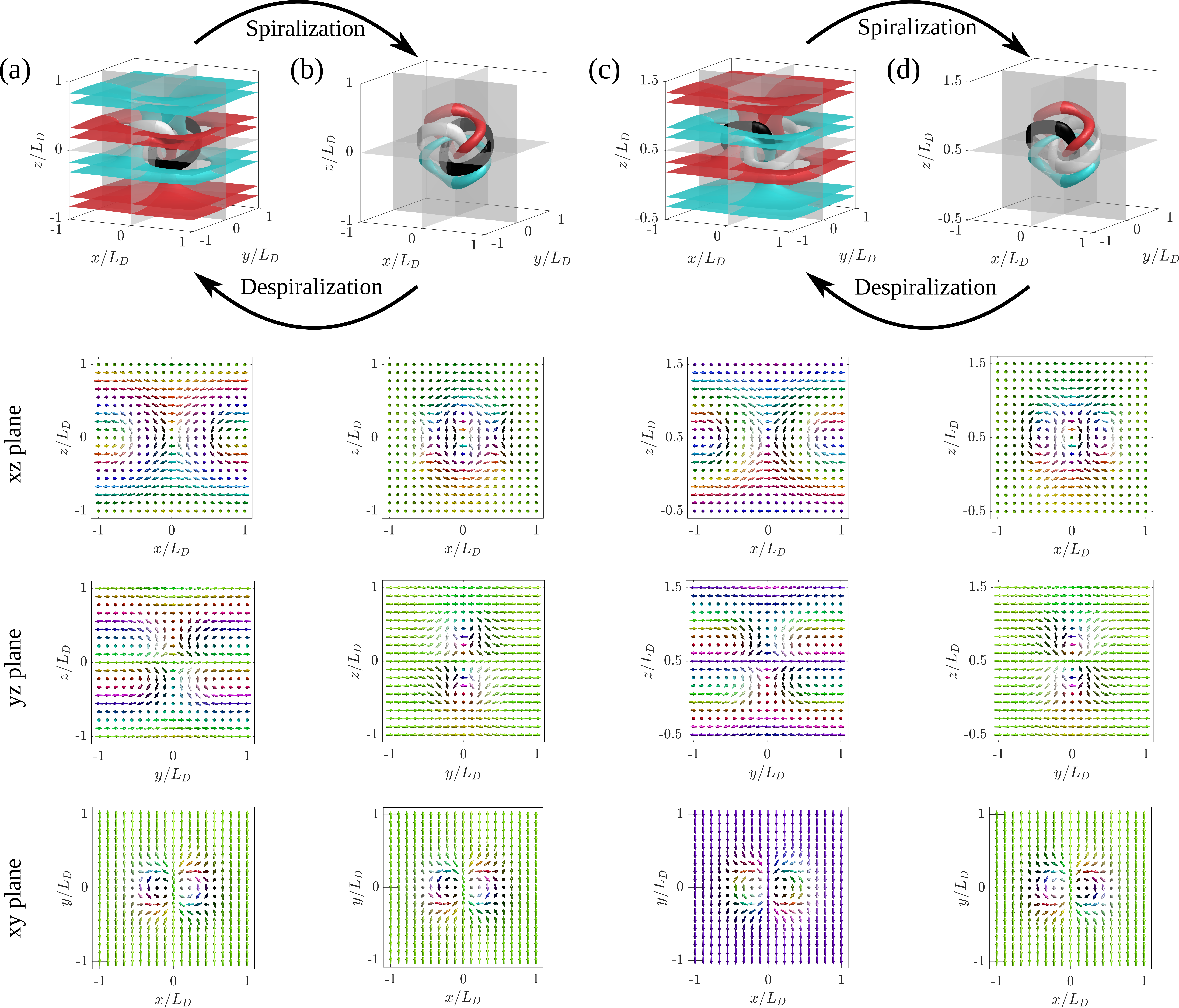}
   \caption{Isosurfaces of $m_z = \pm 0.9$ and $m_x = \pm 0.9$ for 
(a) a heliknoton centered at $z = 0$, 
(b) a Hopfion centered at $z = 0$, 
(c) a heliknoton centered at $0.5\,L_D$, and 
(d) a Hopfion centered at $0.5\,L_D$. 
For each configuration, the corresponding subfigures show cross-sections in the $xz$-, $xy$-, and $yz$-planes.}
    \label{fig:four_isosurface1}
\end{figure*}

For the Hopfion shown in Fig.~\ref{fig:four_isosurface} of the main text, the simulations were initialized using the compact Hopfion ansatz of Ref.~\cite{Knapman2025}, in which the $\mathbf{m} = -\mathbf{e}_y$ isoline forms a circle in the $xz$ plane, see Fig. \ref{fig:Hopfion-ansatz}.

\begin{equation}
\mathbf{m}(x, y, z) = 
\begin{pmatrix}
\sin \Phi \sin \Theta \\
\cos \Theta \\
-\cos \Phi \sin \Theta 
\label{eq:Hopfion_ansatz}
\end{pmatrix},
\end{equation}
where the azimuthal and polar angles are given by
\begin{align}
\Phi &= \psi + \phi, \\
\Theta &= 
\begin{cases}
\pi \exp\!\left( 1 - \dfrac{1}{1 - (\rho / r)^2} \right), & \text{if } \rho < r, \\
0, & \text{if } \rho \ge r.
\end{cases}
\end{align}
The auxiliary radial coordinate $\rho$ and auxiliary angular parameters $\phi$ and $\psi$ that sweep around the minor radius $r$ and major radius $R$ are respectively given by
\begin{subequations}\label{eq:Hopfion_ansatz2}
\begin{align}
\rho &= \sqrt{ y^2 + (x \cos \psi - z \sin \psi - R)^2 }, \label{eq:16a}\\[6pt]
\phi &= \arctan \!\left( \dfrac{ y }{ x \cos \psi - z \sin \psi - R } \right), \label{eq:16b}\\[6pt]
\psi &= \arctan(-z/x). \label{eq:16c}
\end{align}
\end{subequations}
\noindent
Relaxation of this ansatz for $R=0.55 L_D$ and $r=0.45 L_D$ yields a Hopfion centred at $z=0$ embedded in a ferromagnetic background $\vect{m}=\vect{e}_y$, as illustrated in Fig.~\ref{fig:four_isosurface}(a) of the main manuscript.
For a detailed view of different crossections, more isosurfaces, and different heliknoton and Hopfion configurations, see Fig.~\ref{fig:four_isosurface1}. 

\begin{figure}[tb]
    \centering
    \includegraphics[width=1\linewidth]{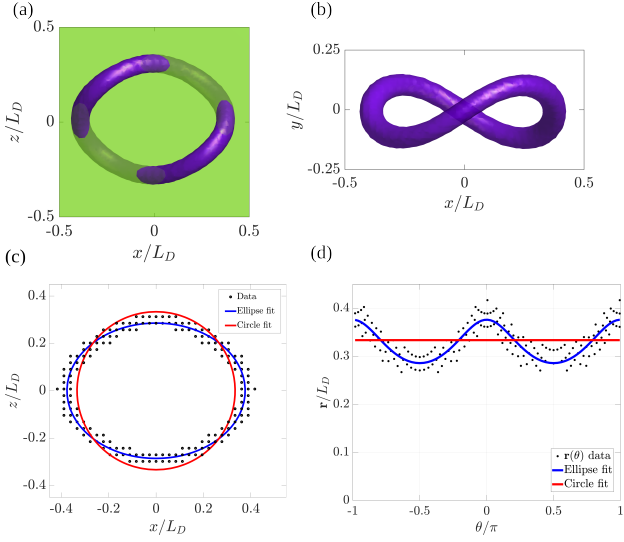}
    \caption{Fourfold distortion of the centreline of the Hopfion, visualized by the $m_y = -0.99$ isosurface (violet tube). (a) The view in the $xz$ plane; the $xz$ plane at $y=0$ is plotted in green for guidance. (b) The view in the $xy$ plane. (c) The $(x,z)$ coordinates of each point contained by the $-0.99$ isosurface (black) together with fits to an ellipse (blue) and a circle (red). (d) The same data shown in polar coordinates $(r,\theta)$, with $r=\sqrt{x^2+z^2}$ and $\theta=\arctan(x/z)$.
    }
    \label{fig:4distortion}
\end{figure}
During relaxation, the initially circular centreline develops an oscillatory out-of-plane displacement with twofold periodicity along the toroidal coordinate and a corresponding fourfold distortion when viewed along the $y$ axis, see Fig.~\ref{fig:4distortion} and discussion below. 

We fit the $(x,z)$-coordinates of the centreline to both a circle and an ellipse. The optimal circular fit yields a radius of 0.35$L_D$. For an elliptical fit of the form $x^2/a^2+y^2/b^2=1/r^2$, the best-fit parameters are $a=0.41L_D$ and $b = 0.29L_D$. These fits are plotted in Fig.~\ref{fig:4distortion}.

\paragraph{Determination of the Hopfion center:}
The center of the Hopfion, $\vect{R}_\mathrm{c} = (x_\mathrm{c},y_\mathrm{c},z_\mathrm{c})$,
is defined as the center of mass of the emergent field $|\mathbf{F}|$,
\begin{equation}
\vect{R}_\mathrm{c} =
\frac{\displaystyle \int \vect{r}\,
|\vect{F}(\vect{r})|\, d^3 r}
{\displaystyle \int |\vect{F}(\vect{r})|\, d^3 r},
\label{eq:Hopfion_center}
\end{equation}
which is a gauge-invariant and numerically robust measure of the position of the toroidal flux tube. 

\paragraph{Determination of the Hopfion center line}
The Hopfion center line is defined relative to the background magnetization. The background magnetization $\mathbf{m}^\mathrm{bg}$
 is obtained by spatially averaging the magnetization field over the simulation volume containing the background, and normalizing the resulting vector. The center line is then identified from the isosurface $|\mathbf{m}(\mathbf{r})-\mathbf{m}^\mathrm{bg}|=2-\delta$. For small $\delta$, this isosurface forms a narrow tubular structure whose medial axis defines the Hopfion center line. This definition is well-suited for Hopfions embedded in a uniform ferromagnetic background; in spatially modulated (spiral) backgrounds, the notion of a unique $\mathbf{m}^\mathrm{bg}$ becomes ambiguous.

Using this procedure, we find that the center line of the relaxed Hopfion is shifted relative to the ansatz in Eq.~\eqref{eq:Hopfion_ansatz} and exhibits a fourfold distortion, as illustrated in Fig.~\ref{fig:4distortion}, where we used $\delta=0.01$.

\paragraph{Determination of the midplane orientation}
\label{sec:midplane}
For the Hopfion configurations considered, the midplane (the green plane in Fig.~\ref{fig:4distortion}) can be described by the plane perpendicular to the normal vector $\mathbf{n}=(\cos\phi,\sin\phi,0)$, where $\phi$ is the azimuthal angle of the Hopfion ring.
To determine $\phi$, we evaluated the weighted quadratic deviation of the
emergent field from the plane,

\begin{equation}
I(\phi) =
\int |\mathbf{F}(\mathbf{r})|\,
\bigl(x\cos\phi + y \sin\phi\bigr)^2
\, d^3 r.
\label{eq:Iphi}
\end{equation}

The midplane corresponds to the orientation that minimizes $I(\phi)$, i.e.,
the plane that divides the toroidal flux distribution most symmetrically with
respect to the $z$-axis.  
In practice, this minimization can be expressed in quadratic form as

\begin{equation}
I(\phi)
= \mathbf{n}^\mathsf{T}
\begin{pmatrix}
\!\!\!\displaystyle\int |\mathbf{F}|\,x^2 d^3r &
\displaystyle\int |\mathbf{F}|\,xy\,d^3r \\[6pt]
\displaystyle\int |\mathbf{F}|\,xy\,d^3r &
\!\!\!\displaystyle\int |\mathbf{F}|\,y^2 d^3r
\end{pmatrix}
\mathbf{n},
\end{equation}

where the symmetric $2\times2$ matrix is the in-plane moment tensor of the emergent magnetic field $|\mathbf{F}|$.
The minimizing direction $\mathbf{n}$ corresponds to the eigenvector of this
tensor with the smallest eigenvalue, yielding the midplane orientation angle

\begin{equation}
\phi = \left[\tan^{-1}\!\left(\frac{n_y}{n_x}\right)\right]
\text{(mod }\pi\text{)}.
\label{eq:phi_angle}
\end{equation}

\begin{figure}
    \centering
    \vspace{0.25cm}
    \includegraphics[width=0.9\linewidth]{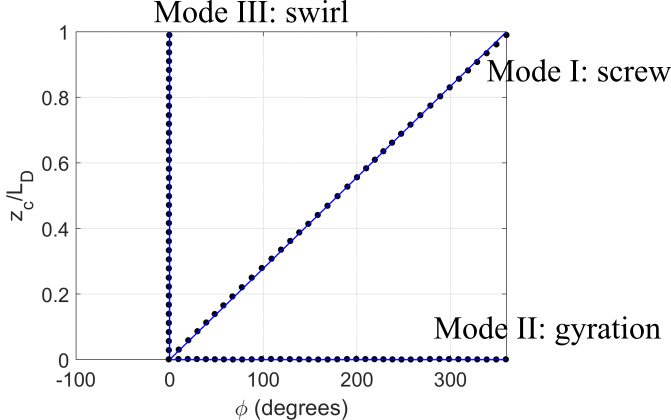}
    \caption{Centre coordinate $z_c$ plotted as a function of the azimuthal angle $\phi$ for three zero modes. The black points represent values calculated using Eqs.~\eqref{eq:Hopfion_center} and \eqref{eq:phi_angle}. The blue lines show the predicted linear behavior of $z_c(\phi)$.}
    \label{fig:threemodes}
\end{figure}

Fig.~\ref{fig:threemodes} shows a plot of the $z$-coordinate of the Hopfion center, $z_c$, as a  function of $\phi$, the orientation of the midplane, obtained using this method, for the three zero modes discussed in Section \ref{sec:Goldstone}.

\section{Other magnetic textures}
\label{sec:othermagn}

\begin{figure}[b]
    \centering
    \includegraphics[width=0.7\linewidth]{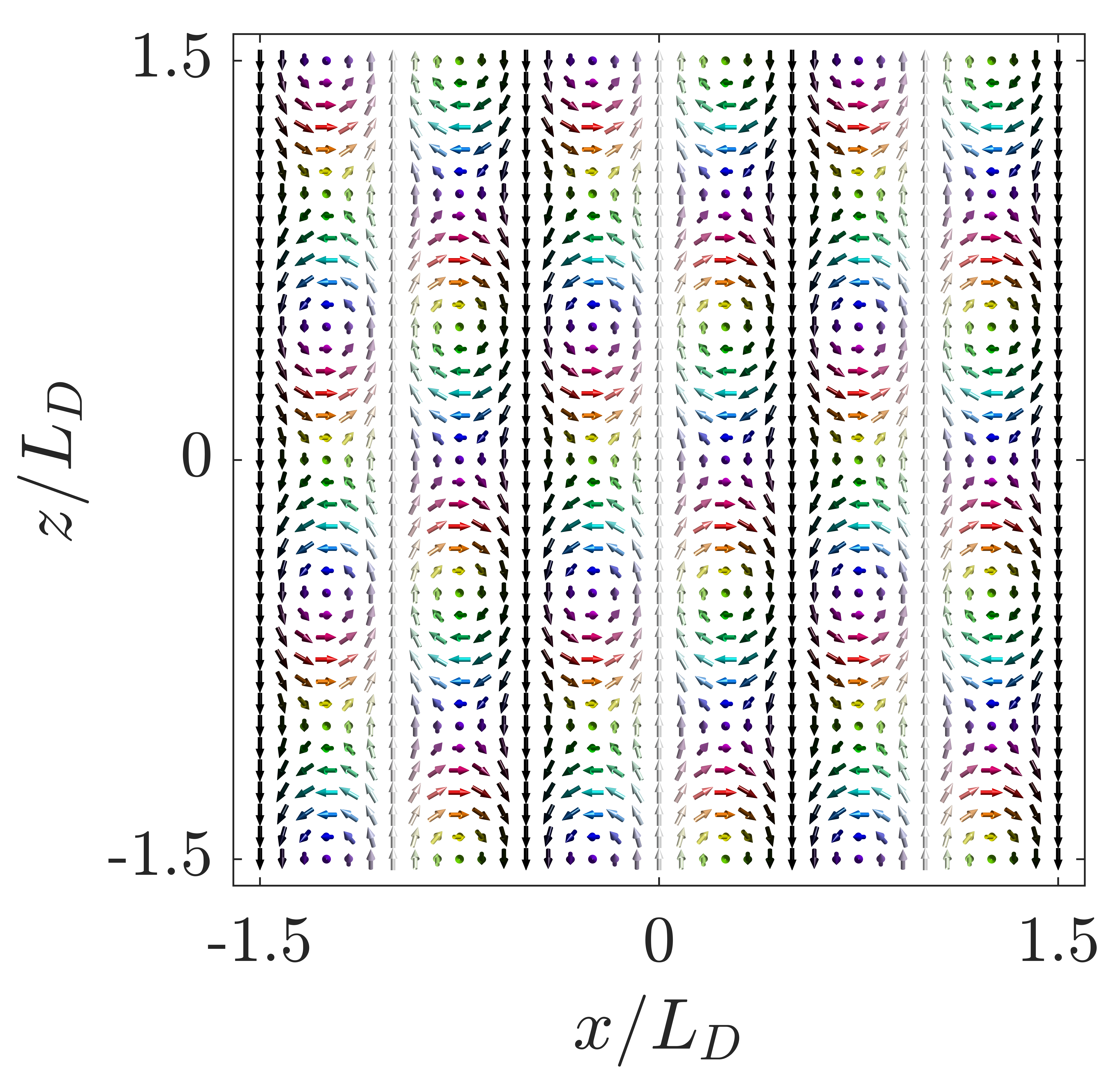}
    \caption{Example of an $xy$ plane cross section of a $2\vect{q}$ spiral state.
    }
    \label{fig:2q spiral}
\end{figure}

For completeness, we present a doubly modulated helical state that arises when the despiralization is performed along a direction different from the original spiral axis. Figure~\ref{fig:2q spiral} shows an \(xz\) cross-section of a representative \(2\vect{q}\) spiral state that is invariant along \(\vect{e}_y\). This texture is obtained by despiralizing a spiral along \(\vect{e}_x\) via a local spin rotation about the \(\vect{e}_z\) axis.

The sinusoidal modulation of the DMI in the screw chiral magnet model along \(z\) produces a continuous interpolation between Bloch- and Néel-type spirals: the texture is purely Néel-type at \(z = n L_D\) and purely Bloch-type at \(z = (n+1)L_D/2\), with integer \(n\). As a result, the state forms an ordered array of Bloch-type vortices and antivortices. Notably, a similar vortex-antivortex pattern appears in the \(xz\) cross-section of the Hopfion shown in the second and fourth columns in Fig.~\ref{fig:four_isosurface1}t, indicating a close structural relation between the two configurations.

\begin{figure*}[t]
    \centering
    \includegraphics[width=1\linewidth]{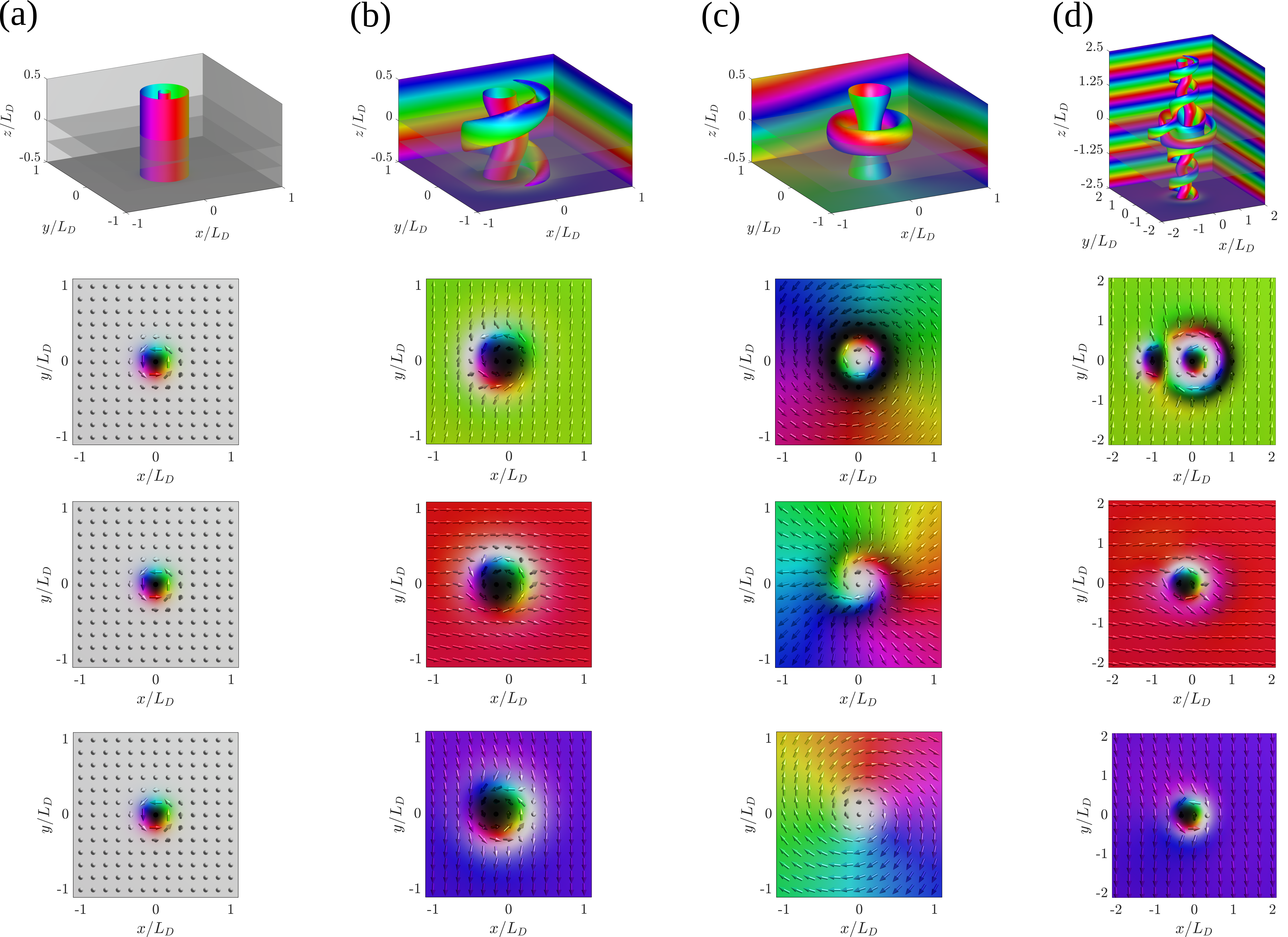}
\caption{
Examples of stable magnetization configurations in the bulk chiral magnet  $\mathcal{E}_{\mathrm{CM}}[\vect{m}] =\: \mathcal{A} \: (\vect{\nabla} \vect{m})^2 \: + \: \mathcal{D} \: \vect{m} \cdot (\vect{\nabla} \times \vect{m})$ that are spiralised versions of those shown in Fig.~\ref{fig:other_textures} of the main text. Highlighted are isosurfaces of $m_z = \pm 0.75$ for the four three-dimensional spin textures: (a) a skyrmion tube in a ferromagnetic background, (b) an in-plane skyrmion tube in a helical spiral background, (c) a Twiston, and (d) a Hopfion configuration surrounding a skyrmion string. For configurations (a)--(c) a full period along the $z$-axis is shown and the subfigures depict $xy$-plane cross-sections at $z = 0$, $z = -L_D/4$, and $z = -L_D/2$. Configuration (d) displays five periods of the skyrmion string with a Hopfion ring located near the $z = 0$ plane. The subfigures for (d) show the $xy$-plane at $z = 0$, $z = -5L_D/4$, and $z = -5L_D/2$.  
}
    \label{fig:other_textures_bcm}
\end{figure*}

Fig.~\ref{fig:other_textures_bcm} shows the simulated textures of the bulk chiral magnet model $\mathcal{E}_{\mathrm{CM}}[\vect{m}] =\: \mathcal{A} \: (\vect{\nabla} \vect{m})^2 \: + \: \mathcal{D} \: \vect{m} \cdot (\vect{\nabla} \times \vect{m})$, corresponding to the pre-despiralized counterparts of Fig.~\ref{fig:other_textures} in the main text.

\end{document}